\documentclass[aip,jap,reprint,10pt,showpacs,unsortaddress,longbibliography]{revtex4-1}
\pdfoutput=1
\usepackage{graphicx}
\usepackage{amsfonts}
\usepackage{amsmath}
\usepackage{natbib}

\begin{document}

%Title of paper
\title{Inter-laboratory nanoamp current comparison with sub-part-per-million uncertainty}

\author{S.~P.~Giblin}
\affiliation{National Physical Laboratory (NPL), Hampton Road, Teddington, Middlesex TW11 0LW, United Kingdom}
\email{stephen.giblin@npl.co.uk}
\author{D. Drung}
\affiliation{Physikalisch-Technische Bundesanstalt (PTB), Abbestra{\ss}e 2-12, 10587 Berlin, Germany}
\author{M. G\"{o}tz}
\author{H. Scherer}
\affiliation{Physikalisch-Technische Bundesanstalt (PTB), Bundesallee 100, 38116 Braunschweig, Germany}

\date{\today}

\begin{abstract}
An ultrastable low-noise current amplifier (ULCA) has been transferred between two laboratories, NPL and PTB, four times, with three of the transfers yielding a relative change in the trans-resistance gain of less than $2 \times 10^{-7}$. This is a new bench-mark for inter-laboratory transfer of small current. We describe in detail the use of a cryogenic current comparator to calibrate the ULCA at NPL, and the use of the ULCA to measure $1$~G$\Omega$ resistors with relative uncertainties at the $10^{-7}$ level.

\end{abstract}

% insert suggested PACS numbers in braces on next line
\pacs{1234}
% insert suggested keywords - APS authors don't need to do this
%\keywords{}

%\maketitle must follow title, authors, abstract, \pacs, and \keywords
\maketitle

\section{\label{IntroSec} Introduction}

Verification of National Metrology Institute (NMI) small-current calibration capabilities by an intercomparison has proved difficult due to the limited stability of commercial ammeters used as travelling standards \cite{willenberg2013euromet}. At the same time, development of prototype single-electron current standards with sub part-per-million (ppm) uncertainty \cite{stein2016robustness,zhao2017thermal} has pushed the limits of existing reference current sources based on, for example, application of a linear voltage ramp to a low-loss capacitor\cite{willenberg2003traceable,van2005accurate,fletcher2007new,callegaro2007current}, and motivated development of the ultrastable low-noise current amplifier (ULCA) \cite{drung2015ultrastable}. The ULCA is a trans-resistance amplifier with 1-year current-to-voltage gain stability of around $1$~$\mu \Omega / \Omega$ under fixed laboratory conditions. It has also demonstrated a gain stability of around $1$~$\mu \Omega / \Omega$ under international transportation from PTB to NPL (UK) and LNE (France) \cite{drung2015validation}, roughly two orders of magnitude better than the instruments used in the intercomparison\cite{willenberg2013euromet}. The LNE and NPL measurements reported in Ref. \cite{drung2015validation} were performed by connecting the ULCA input to a reference current source (NPL) or measurement system (LNE), not by direct calibration of the internal gain elements of the ULCA as performed at PTB \cite{drung2015ultrastable,drung2015improving}, and were limited to an uncertainty of $\sim 1$ part in $10^6$. Here, we report the results of a PTB-NPL comparison\cite{giblin2018interlaboratory} in which both laboratories used cryogenic current comparators (CCCs) to calibrate the ULCA gain directly, with combined uncertainties in both cases less than $0.1$~$\mu \Omega / \Omega$. This low uncertainty allows us to set a new bench-mark for stable transfer of a reference current between two laboratories.

The ULCA converts an input current $I_{\text{IN}}$ to an output voltage $V_{\text{OUT}}$ according to $V_{\text{OUT}} = A_{\text{TR}} I_{\text{IN}}$, where $A_{\text{TR}}$ is the trans-resistance gain nominally equal to $1$~G$\Omega$. $A_{\text{TR}}$ is composed of two components which can be calibrated separately: the dimensionless current gain $G_{\text{I}}$ and the trans-resistance $R_{\text{IV}}$. $G_{\text{I}}$ and $R_{\text{IV}}$ are nominally equal to $1000$ and $1$~M$\Omega$ respectively, and $A_{\text{TR}} = G_{\text{I}} R_{\text{IV}}$. The calibration of the ULCA at PTB using a CCC has already been described in detail \cite{drung2015ultrastable,drung2015improving}. In this paper, we describe the calibration of the ULCA at NPL using the NPL high-resistance CCC \cite{fletcher2000cryogenic}, which differs in design from the CCC used at PTB. We then present the results of the comparison, in which the ULCA was transferred twice from PTB to NPL and back. Finally, we investigate the use of the ULCA at NPL to calibrate $1$~G$\Omega$ resistors, and compare it to the existing calibration method using a CCC. The bulk of the measurements reported in this paper were performed on the `transfer ULCA'. However, we also report some NPL measurements on another ULCA unit, the `NPL ULCA', purchased commercially from Magnicon and delivered in February 2018.

A note on nomenclature: in describing deviations of a measured quantity from a nominal value, we employ the following notation: the quantity $X$ with nominal value $X_{\text{NOM}}$ can be written $X = X_{\text{NOM}} (1 + \delta X)$, where $\delta X$ is the deviation from nominal. Following common metrological practice for cases where $\delta X \ll 1$, all numerical graph labels in this paper show $\delta X$ in parts in $10^6$. All uncertainties are standard uncertainties $(k=1)$ unless stated otherwise.

\begin{figure}
\includegraphics[width=8.5cm]{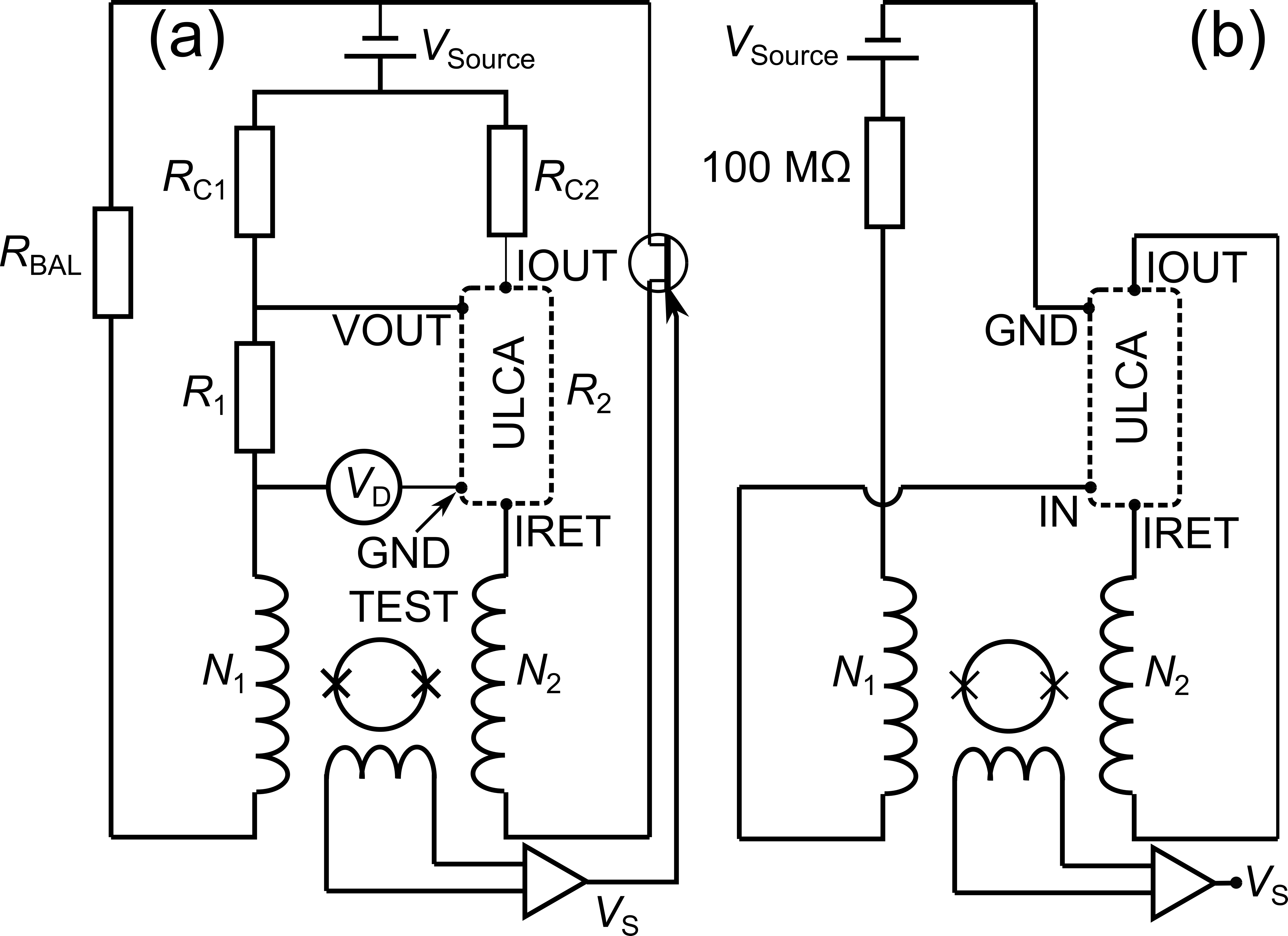}
\caption{\label{CircuitFig}\textsf{Schematic circuit diagrams showing the calibration of (a) $R_{\text{IV}}$ and (b) $G_{\text{I}}$ using the NPL high-resistance CCC. In both diagrams the ULCA unit is indicated by a dashed box. The ULCA settings are input = SRC, output = VOUT in (a), and input = AMP, output = IOUT in (b).}}
\end{figure}

\section{\label{RIVSec} Calibration of ULCA $R_{\text{IV}}$ at NPL}

Figure \ref{CircuitFig} (a) shows a schematic diagram for the calibration of $R_{\text{IV}}$ using the NPL high-resistance CCC. In this diagram, as elsewhere in this paper, the ULCA terminals are labeled IOUT, VOUT, etc following Ref. \cite{drung2015ultrastable}. The bridge was set up in the same way as for calibration of a 4-terminal standard resistor $R_{\text{2}}$ of nominal value $1$~M$\Omega$, in terms of a known standard $R_{\text{1}}$ with nominal value $100$~k$\Omega$, with $N_{\text{1}} = 1000$ and $N_{\text{2}} = 10000$. Some measurements were also performed with $R_{\text{1}} =1$~M$\Omega$, and $N_{\text{1}} = N_{\text{2}} = 10000$. In this CCC, a single voltage source energizes both arms of the bridge, and the SQUID output voltage $V_{\text{S}}$ is used as the input to a servo which controls the resistance of a JFET in one of the arms. A combining network composed of resistors $R_{\text{C1}}$ and $R_{\text{C2}}$, nominally equal to $R_{\text{1}}$ and $R_{\text{2}}$ respectively, ensures that current flowing in the high-potential voltage terminals of the resistors does not affect the measurement. The raw bridge data consists of readings of the detector voltage $V_{\text{D}}$, and measurements are performed in two phases, with an extra turn added to $N_{\text{2}}$ in the second phase to calibrate the overall CCC gain. The change in the detector voltage when the bridge excitation is reversed is denoted $\Delta V_{\text{D}}$ with $N_{\text{2}} = 10000$, and $\Delta V_{\text{D}}'$ with $N_{\text{2}}' = 10001$. The ratio of the two resistances is then given by 

\begin{equation}
\frac{R_{\text{2}}}{R_{\text{1}}} = \frac { \frac{N_{\text{2}}}{n} - v_{\text{D}} \frac{N_{\text{2}}'}{n'}}     { \frac{N_{\text{1}}}{n} - v_{\text{D}} \frac{N_{\text{1}}}{n'}}.
\end{equation}

with $n = N_{\text{1}} + N_{\text{2}}$, $n' = N_{\text{1}} + N_{\text{2}}'$ and $v_{\text{D}} = \Delta V_{\text{D}} / \Delta V_{\text{D}}'$

\begin{figure}
\includegraphics[width=8.5cm]{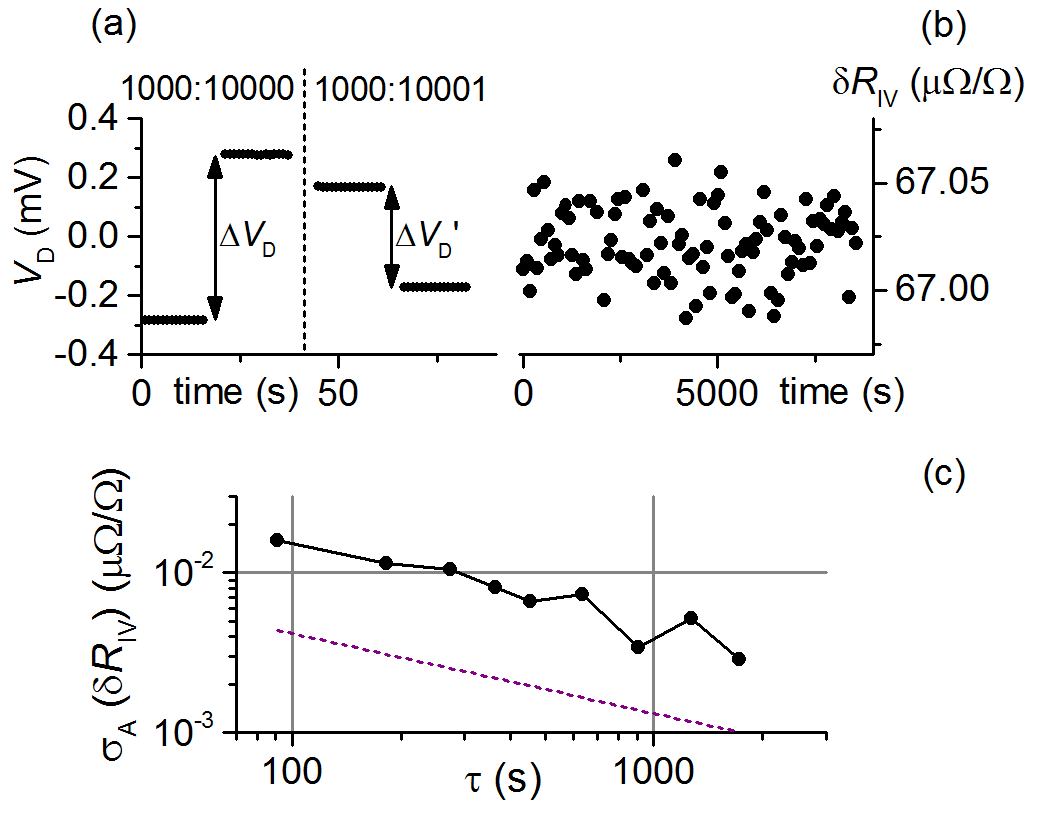}
\caption{\label{RIVFig}\textsf{(a): Raw detector data for a calibration of $R_{\text{IV}}$ using the NPL CCC, for one calibration cycle. (b) values of $\delta R_{\text{IV}}$ calculated from a calibration, where $R_{\text{IV}} = 1$~M$\Omega (1+\delta R_{\text{IV}})$. Each data point is calculated from one cycle of the type shown in (a). (c): Allan deviation of the data in (b). The dashed line with gradient $1 / \sqrt{\tau}$ is the expected Allan deviation due to the noise of the ULCA trans-resistance gain stage.}}
\end{figure}

Raw bridge detector data for a single calibration cycle is shown in figure \ref{RIVFig} (a), illustrating the two difference signals $\Delta V_{\text{D}}$ and $\Delta V_{\text{D}}'$. The bridge voltage source was set to $\pm 9$~V, so that the $5$~V limit of the ULCA output stage was not exceeded (half of the source voltage is dropped across the combining network resistors). Figure \ref{RIVFig} (b) shows values of $\delta R_{\text{IV}}$, the dimensionless deviation of $R_{\text{IV}}$ from its nominal $1$~M$\Omega$ in parts in $10^6$, calculated from $91$ cycles of the type shown in panel (a). The Allan deviation of these values is plotted in panel (c), showing that a relative type A uncertainty of less than $10^{-8}$ is reached after only a few minutes. The noise contribution of the ULCA is $\approx 160$~fA$/ \sqrt{\text{Hz}}$, dominated by the $\approx 130$~fA$/ \sqrt{\text{Hz}}$ thermal noise in the $1$~M$\Omega$ ULCA output resistor \cite{drung2015ultrastable}. The Allan deviation due to the ULCA noise is shown as a dashed line in figure \ref{RIVFig} (c), and it is clear that there is a large amount of extra noise present. The bridge detector voltage noise and the CCC SQUID noise contributions are both roughly half as large as the ULCA noise. The origin of total measurement noise is not currently understood, and may be due to pickup of interference. The dominant component of the relative combined uncertainty in $R_{\text{IV}}$ is the $4 \times 10^{-8}$ type B standard uncertainty in the $100$~k$\Omega$ reference resistor. For calibrations with a $1:1$ ratio, the $1$~M$\Omega$ reference resistor contributed a relative type B uncertainty of $6 \times 10^{-8}$. In practice, to ensure SQUID stability, the ULCA battery box / charger unit had to be unplugged from the mains during the $R_{\text{IV}}$ calibrations, and the battery box earth connector (corresponding to the battery charger ground) connected to the CCC electronics ground. The ULCA temperature was not recorded continuously during calibrations. It was recorded before and after the calibration, and the mean of these two values reported as the calibration temperature.

\section{\label{GISec} Calibration of ULCA $G_{\text{I}}$ at NPL}

The calibration of the dimensionless current gain $G_{\text{I}}$ required some modification of the CCC bridge circuit, as shown in figure \ref{CircuitFig} (b). A $100$~M$\Omega$ resistor in series with the bridge voltage source generated a current of $\sim \pm 13$~nA \footnote{Precise adjustment of the bridge voltage source, which has a maximum output of $100$~V, proved difficult, and the current was not therefore precisely controlled from one calibration to the next. It varied between about $\pm 11$~nA and $\pm 13$~nA}, which flowed in a winding of $N_{\text{1}} = 10000$ turns, and into the ULCA input. The second stage of the ULCA was then used to drive a current in a second winding with $N_{\text{2}} = 10$ turns. The output of the SQUID control electronics, $V_{\text{S}}$, was recorded, and the difference signal $\Delta V_{\text{S}}$ extracted when the excitation current was reversed. As in the case of the $R_{\text{IV}}$ calibration, an additional measurement was performed with $N_{\text{1}}' = 10001$ turns, yielding a difference signal $\Delta V_{\text{S}}'$. If we define the deviation of the current gain from its nominal value such that

\begin{equation}
G_{\text{I}} = \frac{I_{\text{2}}}{I_{\text{1}}} = 1000 \big(1 + \delta G_{\text{I}} \big),
\end{equation}

then the current gain can be calculated from the CCC data as:

\begin{equation}
\delta G_{\text{I}} = \frac{ v_{\text{S}}(N_{\text{1}}' - N_{\text{1}})}        {N_{\text{1}} (v_{\text{S}}-1) }
\end{equation}

with $v_{\text{S}}=\Delta V_{\text{S}}/\Delta V_{\text{S}}'$. By co-incidence, the ULCA used for the comparison had $G_{\text{I}}$ very close to its nominal value, $|\delta G_{\text{I}}|<10^{-6}$, and consequently $\Delta V_{\text{S}} \ll \Delta V_{\text{S}}'$. Using this approximation and entering the values for the numbers of turns into equation (3), we obtain the simple relation

\begin{equation}
\delta G_{\text{I}} \approx \frac{- v_{\text{S}}}{10^{4}}.
\end{equation}

\begin{figure}
\includegraphics[width=8.5cm]{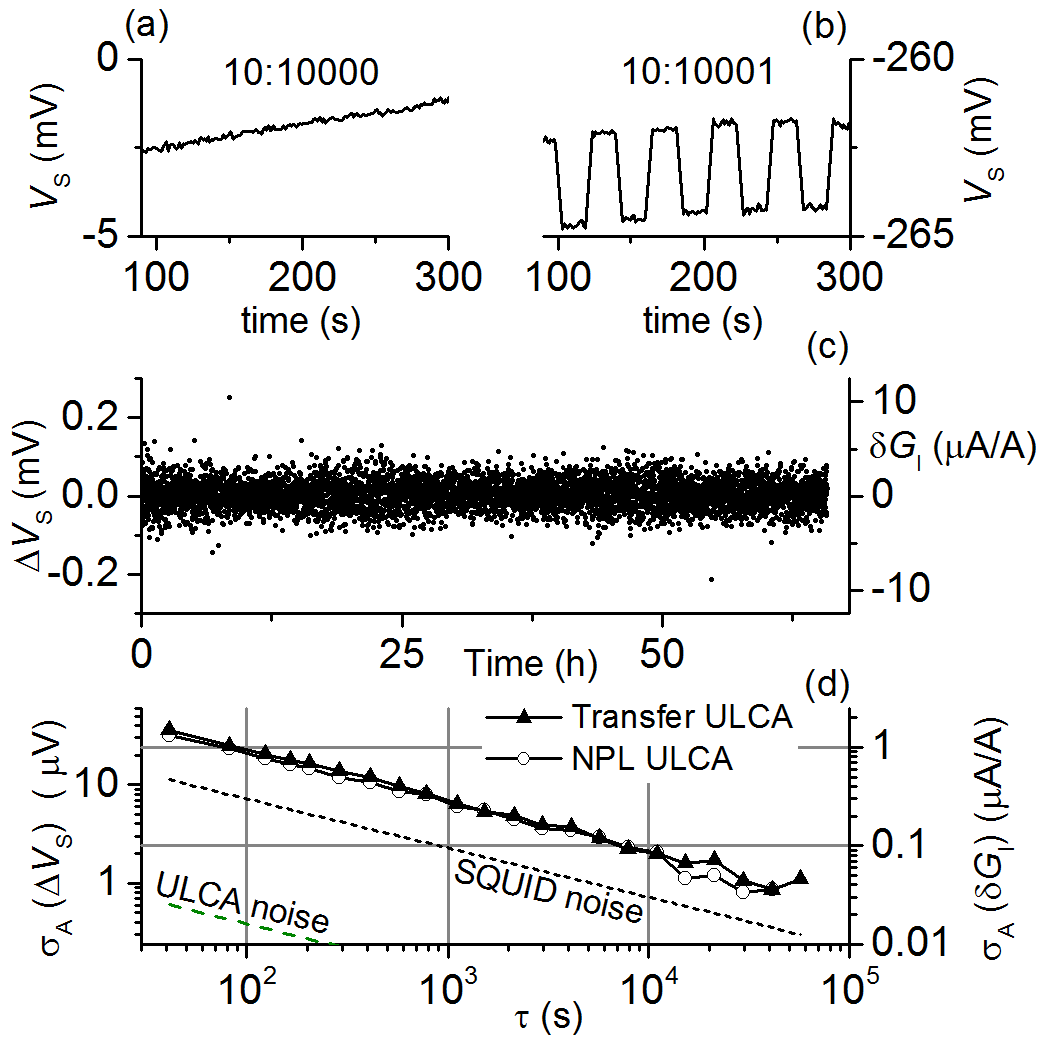}
\caption{\label{GIFig}\textsf{(a): Short section of raw SQUID data from a calibration of $G_{\text{I}}$. (b): as (a), but with an additional turn added to the large winding. (c): SQUID difference voltage extracted from the raw calibration data (left axis) and the difference voltage scaled using equation (4) to the deviation of $G_{\text{I}}$ from its nominal value in parts in $10^6$ (right axis):  $G_{\text{I}} = 10^{3}(1 + \delta G_{\text{I}})$ (d): Filled triangles: Allan deviation of the data in (c). Open circles: Allan deviation of another calibration data set on the NPL ULCA. The left and right axes have the same meaning as in plot (c). The dashed lines with gradient $1/ \sqrt{\tau}$ are the Allan deviations expected from (upper line): the SQUID noise measured in separate experiments with all electronics disconnected from the CCC windings and (lower line): the ULCA input current noise flowing in the $10^4$-turn CCC winding.}}
\end{figure}

The calibration of $G_{\text{I}}$ is illustrated in figure \ref{GIFig}. In panels (a) and (b) we show short sections of raw SQUID data from forward-reverse cycles with $10:10000$, and $10:10001$ turns respectively. Due to $G_{\text{I}}$ being very close to its nominal value, the difference signal is not visible in panel (a), but $\Delta V_{\text{S}}' \sim 2.5$~mV is clearly visible in panel (b). A useful consequence of the very small $\delta G_{\text{I}}$ is that the measurement of $\Delta V_{\text{S}}'$ contributes only a small correction to $\delta G_{\text{I}}$ calculated using equation (4), and therefore it is not necessary to evaluate $\Delta V_{\text{S}}'$ with small uncertainty. In practice, a long measurement of $\Delta V_{\text{S}}$, consisting of $\sim 5000$ forward-reverse cycles (typically performed over a weekend) was preceded and followed by short measurements of $\Delta V_{\text{S}}'$ of $\sim 10$ cycles each, with the average value of $\Delta V_{\text{S}}'$ being used in equation (4). To remove errors due to linear drift in $V_{\text{S}}$, the difference signals $\Delta V_{\text{S}}$ and $\Delta V_{\text{S}}'$ were extracted from the raw data using a standard algorithm which evaluates the difference using a `reverse' data segment, and the two adjacent `forward' half-segments. This algorithm is explained in more detail in the supplementary information to Ref. \cite{giblin2017robust}. 

Figure \ref{GIFig} (c) shows values of $\Delta V_{\text{S}}$ extracted from a typical long measurement (of which panel (a) shows a small segment), and panel (d), filled triangles, shows the Allan deviation of this data, both as SQUID difference voltage (left axes) and $\delta G_{\text{I}}$ (right axes, after scaling using equation (4)). For comparison, we show the Allan deviation expected from the SQUID noise of $\approx 7.5 \times 10^{-5} \Phi_{0} / \sqrt{\text{Hz}}$ measured in separate experiments with the CCC windings disconnected from the ULCA and the CCC electronics, and scaled to allow for the duty cycle of the $G_{\text{I}}$ calibration, in which $\frac{1}{3}$ of the data was discarded to reject transient effects following current reversal. We also show the contribution of the ULCA input current noise of $\approx 2.4$~fA$/ \sqrt{\text{Hz}}$, which is a negligible contribution for our setup. The Allan deviation of the $G_{\text{I}}$ calibration data is roughly a factor $3$ higher than expected from the SQUID noise alone. The reason for this increase is not known, but we note that the $G_{\text{I}}$ calibration at NPL does not use feedback to maintain CCC flux balance, unlike the equivalent calibration at PTB. Figure \ref{GIFig} (d) can be compared with figure 12 of Ref. \cite{drung2015ultrastable}, and it will be immediately apparent that calibration of $G_{\text{I}}$ at NPL is more than a factor $20$ noisier than at PTB using the 14-bit CCC \cite{drung2015ultrastable,drung2015improving}. At PTB, an Allan deviation of $\sigma _{\text{A}} (\delta G_{\text{I}}) = 1 \times 10^{-8}$ is reached after $\sim 2000$~seconds, whereas this level of uncertainty in principle could be attained after several days with the NPL CCC, and was not in fact reached in any of the $G_{\text{I}}$ calibrations performed to date. However, we note that the NPL CCC was originally designed for routine calibrations of high-value standard resistors \cite{fletcher2000cryogenic} at considerably higher currents than those employed here in the calibration of $G_{\text{I}}$. There is no indication of a transition to frequency-dependent noise in the Allan deviation plots of figure \ref{GIFig} (d), or in any of the other $G_{\text{I}}$ calibration runs, and we evaluated the type A uncertainty as the standard error of the mean. For the calibration run presented in the figure, $\delta G_{\text{I}} = (-0.296 \pm 0.021) \times 10^{-6}$, where the relative uncertainty includes a $1 \times 10^{-8}$ type B component. In figure \ref{GIFig} (d) we also show the Allan deviation of a later (May 2018) calibration of the NPL ULCA (open circles). This shows that the noise level is a stable property of the CCC.

Overall, it was possible to calibrate the ULCA at NPL with a relative combined uncertainty in $A_{\text{TR}}$ less than $10^{-7}$. The main components of the relative uncertainty were the $4 \times 10^{-8}$ type B uncertainty in the $100$~k$\Omega$ reference resistor used to calibrate $R_{\text{IV}}$, and the few parts in $10^{-8}$ type A uncertainty in $G_{\text{I}}$. As noted, the NPL CCC was not in any way specialized towards ULCA calibration, and for widespread adoption of the ULCA it is encouraging that such satisfactory results could be obtained with this CCC.

\section{\label{MainSec}Results of the comparison}

\begin{figure}
\includegraphics[width=8.5cm]{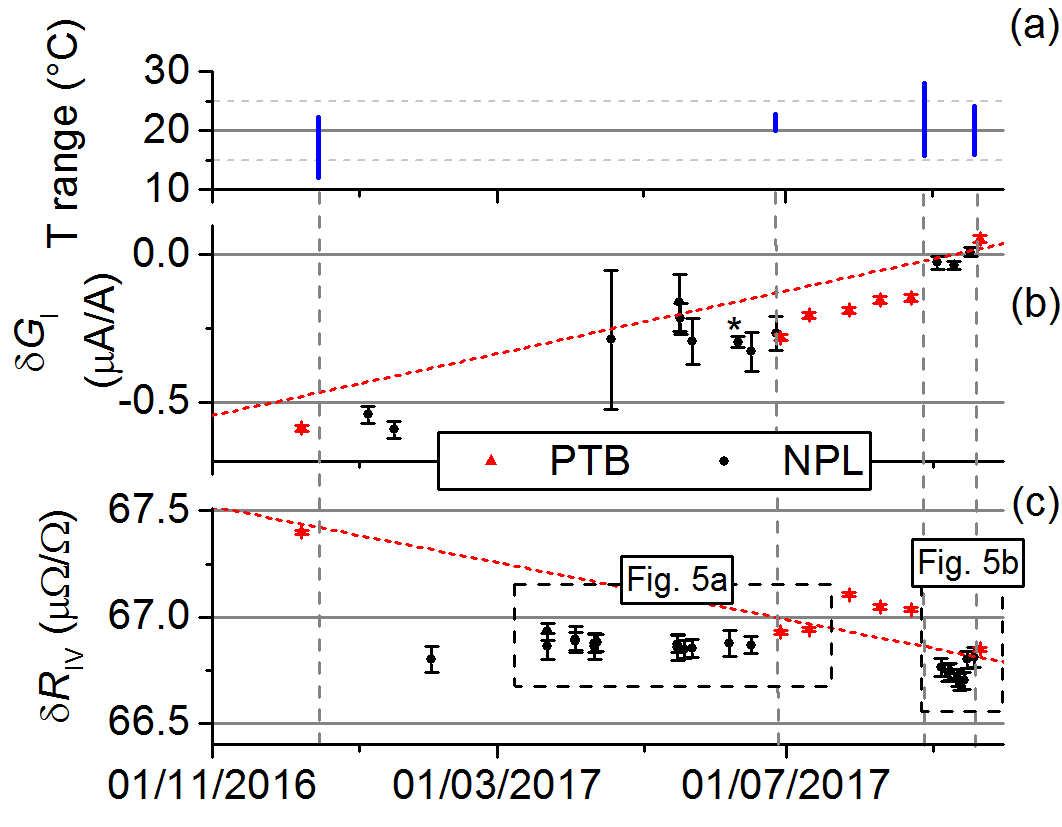}
\caption{\label{DriftFig}\textsf{Calibrations of (b): $G_{\text{I}}$ and (c): $R_{\text{IV}}$ at NPL and PTB. Vertical dashed lines: dates of ULCA transfers between laboratories. Dotted lines: linear fits to previous calibrations at PTB. $G_{\text{I}} = 1000(1+\delta G_{\text{I}})$ and $R_{\text{IV}} = 1$~M$\Omega (1+\delta R_{\text{IV}})$. All error bars indicate combined uncertainty. (a): temperature range logged during the ULCA transfers.}}
\end{figure}

The transfer ULCA spent two periods of time at NPL: 16th December 2016 - 26th June 2017, and a shorter period from 31st August - 18th September 2017. All four transfers (PTB to NPL and back, twice) were done by commercial courier with transport times of $\lesssim 48$~hours, and the ULCA package was accompanied on all transfers by a temperature logger with $0.1^{\circ}$C resolution. We now present the results of the comparison, in which the ULCA was calibrated at PTB and NPL. All values of $\delta G_{\text{I}}$ and $\delta R_{\text{IV}}$ are corrected to a standard temperature using logs of the ULCA TEMP output, and the temperature coefficients of  $-0.065 \times 10^{-6} / ^\circ$C and $-0.101 \times 10^{-6} / ^\circ$C respectively, which were measured in preliminary characterisation at PTB. Values for $\delta G_{\text{I}}$ and $\delta R_{\text{IV}}$ measured at both laboratories, are shown in figures \ref{DriftFig} (b) and (c), with the temperature range logged during the ULCA transfers shown as vertical bars in panel (a). The dates of the four transfers are indicated by vertical dashed lines. The ULCA was calibrated at PTB five times over a period of $1.5$ years before the first transfer. Only the last of these calibrations are shown in figure \ref{DriftFig} (b,c), but linear least-squares fit to all five calibrations are shown as dashed lines. These fits established the long-term drift of the ULCA as $d (\delta R_{\text{IV}}) / dt = -0.80 \times 10^{-6}$/year and  $d (\delta G_{\text{I}}) / dt = 0.64 \times 10^{-6}$/year. Broadly, the values of $\delta G_{\text{I}}$ measured at both laboratories follow this long-term drift line, but $\delta R_{\text{IV}}$ exhibited a $\sim 0.5 \times 10^{-6}$ discontinuity after the first transfer, remained roughly time-independent during the first period at NPL, and recovered the drift line towards the end of the measurement period. 

\begin{figure}
\includegraphics[width=8.5cm]{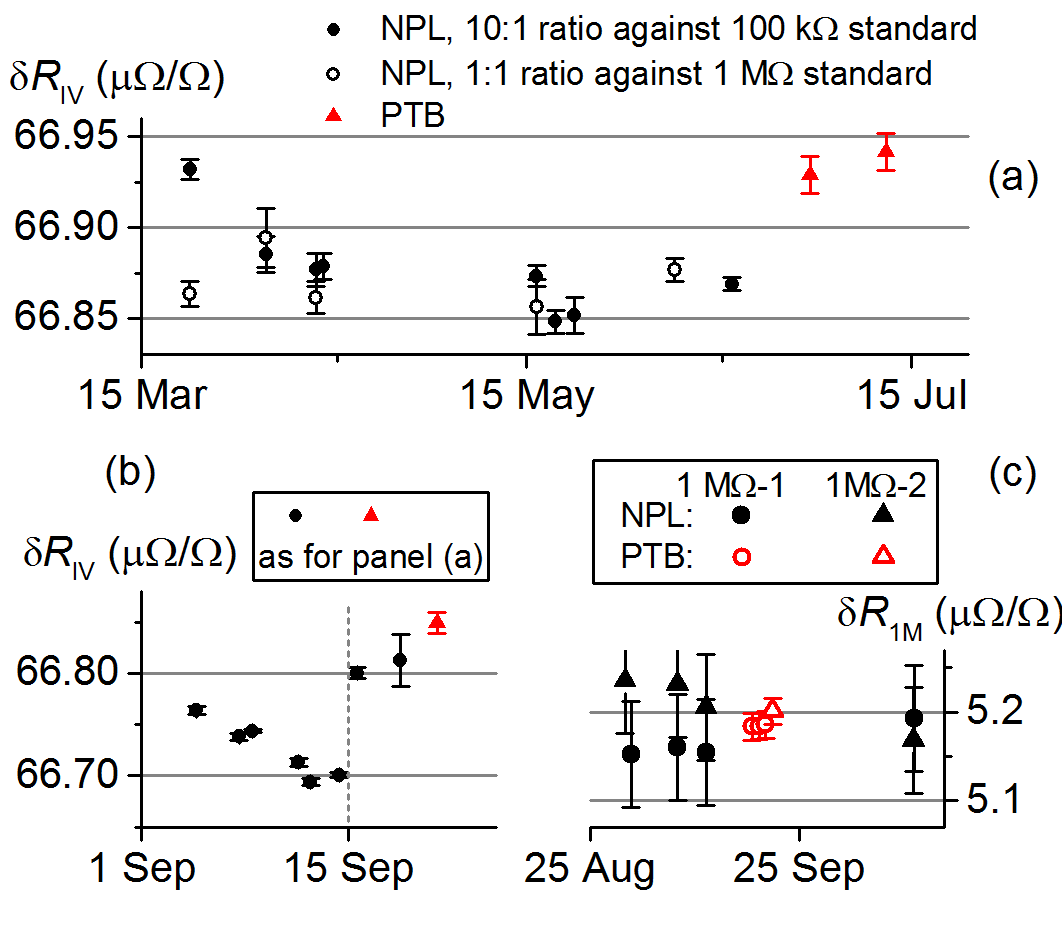}
\caption{\label{ZoomDriftFig}\textsf{(a) and (b): Portions of the $R_{\text{IV}}$ calibration data from figure \ref{DriftFig} on an expanded time axis. Vertical dashed line in (b) indicates internal transport of the ULCA to a different laboratory at NPL. Error bars indicate combined uncertainty for PTB, and type A uncertainty for NPL. (c): Results of the $1$~M$\Omega$ comparison. $R_{\text{1M}} = 1$~M$\Omega (1+\delta R_{\text{1M}})$. Error bars indicate combined uncertainty. All x-axis date labels refer to the year 2017.}}
\end{figure}

Figures \ref{ZoomDriftFig} (a,b) show two portions of the figure \ref{DriftFig} (b) on expanded axes. These figures distinguish between NPL calibrations of $R_{\text{IV}}$ against $100$~k$\Omega$ or $1$~M$\Omega$ reference standards. When both types of calibration were done on the same day, the agreement in $\delta R_{\text{IV}}$ was a few parts in $10^8$, apart from the first pair, where the cause of the $7 \times 10^{-8}$ difference is not known. Figure \ref{ZoomDriftFig} (b) also shows an interesting transport effect: The ULCA was transported under battery power to a different laboratory at NPL for 1 day, then moved back to the calibration laboratory. The date of this transport is indicated by a vertical dashed line, and there is a clear $1 \times 10^{-7}$ shift in $\delta R_{\text{IV}}$. In comparison, $\delta G_{\text{I}}$ changed by only $4 \times 10^{-8}$. Altogether, the data on transporting this ULCA unit suggest that $R_{\text{IV}}$ is more sensitive to transport effects than $G_{\text{I}}$.

\begin{table*}
\caption{\label{SummaryTable}Summary of ULCA transfer stability. Transfers 1-4 were, in chronological order, performed on the transfer ULCA on dates indicated by vertical dashed lines in figures \ref{DriftFig} (b) and (c). Transfer 5 was performed on the NPL ULCA in January 2018. The changes in the three parameters $G_{\text{I}}$, $R_{\text{IV}}$ and $A_{\text{TR}}$ are dimensionless differences in parts in $10^{6}$.}
\centering
\setlength{\tabcolsep}{8pt}
\begin{tabular}{c c c c c c}
\hline\hline
Transfer no. & $T_{\text{MAX}}$ ($^{\circ}$C) & $T_{\text{MIN}}$ ($^{\circ}$C) & $\Delta G_{\text{I}} (\mu \text{A} / \text{A})$ & $\Delta R_{\text{IV}} (\mu \Omega / \Omega)$ & $\Delta A_{\text{TR}} (\mu \text{A} / \text{A})$ \\[0.5ex]
\hline
$1$ & $22.3$ & $12.1$ & $0.046 \pm 0.032$ & $-0.598 \pm0.062$ & $-0.552 \pm 0.069$ \\
$2$ & $22.8$ & $20.1$ & $-0.013 \pm 0.058$ & $0.060 \pm 0.041$ & $0.047 \pm 0.071$ \\
$3$ & $28.0$ & $15.8$ & $0.120 \pm 0.024$ & $-0.272 \pm 0.041$ & $-0.152 \pm 0.047$ \\
$4$ & $24.1$ & $16.0$ & $0.044 \pm 0.020$ & $0.036 \pm 0.048$ & $0.080 \pm 0.052$ \\
$5$ & - & - & $-0.129 \pm 0.020$ & $-0.505 \pm 0.061 $ & $-0.634 \pm 0.064$ \\
\hline
\end{tabular}
\end{table*}

The key finding of this study concerns the stability of the ULCA under international transportation, and these results are summarised in table \ref{SummaryTable}. In addition to the four journeys of the transfer ULCA (rows 1-4 in the table), we also show, as row 5 of the table, the parameters for the NPL ULCA, calibrated at PTB in January 2018 and subsequently measured at NPL in February 2018. For this transfer, the temperature of the ULCA package was not recorded. The last three columns show the differences in $G_{\text{I}}$, $R_{\text{IV}}$ and $A_{\text{TR}}$, defined in each case as the value before transportation subtracted from the value after transportation and expressed as parts in $10^{6}$ of the nominal value. Three of the transfers yielded a relative change in $A_{\text{TR}}$ of less than $2 \times 10^{-7}$. The first transfer of the comparison yielded an exceptional change, in which $A_{\text{TR}}$ jumped down by $5 \times 10^{-7}$, almost entirely due to the jump in $R_{\text{IV}}$ clearly visible in figure \ref{DriftFig} (c). However, $R_{\text{IV}}$ was measured at NPL roughly 1.5 months after the transfer took place, whereas for all the other transfers, measurements were made within just a few days of the transfer. It is possible that the jump in $R_{\text{IV}}$ solely due to the first transportation is exaggerated, and some of the change took place after the transfer to NPL. However, the roughly constant $R_{\text{IV}}$ at NPL over the subsequent 4.5 months suggest that this is not the case, and that the shift indeed took place during the transportation. The minimum temperature logged during transfer 1 was lower than during transfers 2-4, and in this context the data for transfer 5 is noteworthy. Although the temperature was not recorded, transfer 5 also took place during winter time, and a similar (both in magnitude and sign) shift in $R_{\text{IV}}$ was measured. Measurements of $R_{\text{IV}}$ of this ULCA at NPL up to May 2018 showed a relative drift from the first post-transfer value of at most $1$ part in $10^{7}$.

\section{$1$~M$\Omega$ comparison}

Because the ULCA $R_{\text{IV}}$ is calibrated as a $1$~M$\Omega$ resistor, we augmented our comparison of the ULCA with a bi-lateral intercomparison of $1$~M$\Omega$ standard resistors to establish a base-line of agreement with our respective measurement systems. Two resistors maintained at NPL, both of type Fluke 742A and co-incidentally very close in value, were transferred to PTB and back to NPL in September 2017. They were measured in both laboratories as per standard calibration procedures. The results of the comparison are shown in figure \ref{ZoomDriftFig} (c). The agreement between the two laboratories is better than $5 \times 10^{-8}$ and well within the uncertainties. This result provides added confidence that changes in $R_{\text{IV}}$ observed in the previous section can be attributed to the behavior of the ULCA, and not to any systematic errors in the resistance traceability at NPL or PTB.

\section{\label{1GSec} Calibration of $1$~G$\Omega$ resistors at NPL using the ULCA}

In this final section, we explore the use of the ULCA as a route to calibrating $1$~G$\Omega$ resistors at NPL. The same CCC which was used to calibrate $G_{\text{I}}$ and $R_{\text{IV}}$ is in routine use for calibration of decade standard resistors from $100$~k$\Omega$ up to $1$~G$\Omega$, with a calibration and measurement capability (CMC) relative uncertainty ($k=2$) of $1.6 \times 10^{-6}$ for $1$~G$\Omega$. In addition to routine commercial calibrations, this traceability route to $1$~G$\Omega$ has played a key role in research into single-electron current sources.  Five major measurement campaigns on electron pumps at NPL \cite{giblin2012towards,bae2015precision,yamahata2016gigahertz,giblin2017robust,zhao2017thermal} have used a reference current source composed of a temperature-controlled $1$~G$\Omega$ standard resistor (Guildline type 9336), and calibrated voltmeter. Here, we used the ULCA to calibrate the same resistor, known as the `SET $1$~G$\Omega$', and compared these calibrations with calibrations made using the established CCC route.

\begin{figure}
\includegraphics[width=8.5cm]{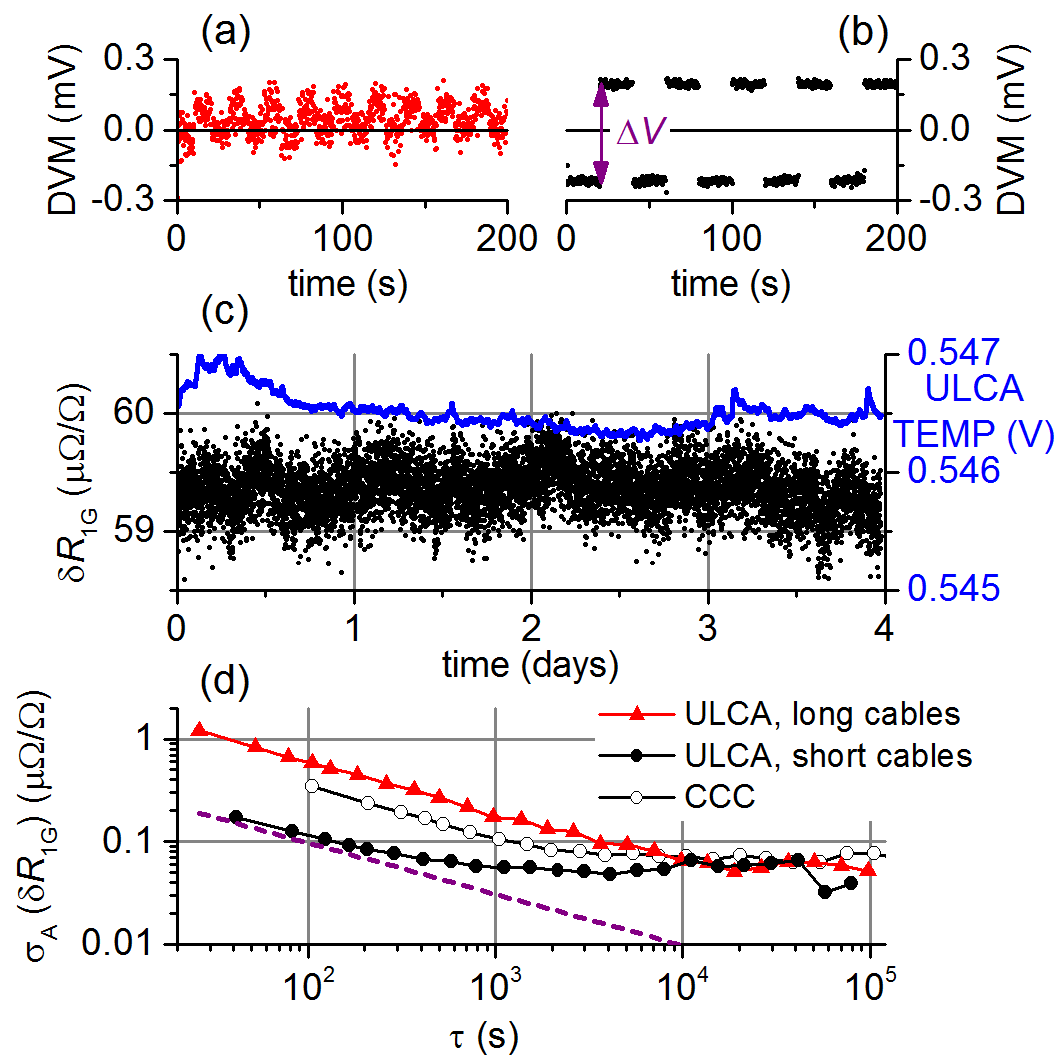}
\caption{\label{1GCalFig}\textsf{(a): section of raw DVM data from a calibration of a $1$~G$\Omega$ using the transfer ULCA. The resistor and ULCA were sited in adjacent laboratories and connected by cables $\sim 10$~m long. (b): as (a), but using a different ULCA unit. The same $1$~G$\Omega$ resistor is sited next to the ULCA and connected with short cables $\sim 1$~m long. The vertical double arrow indicates the difference voltage $\Delta V$ extracted from the raw data. (c): (points, left axis): values of the resistor calculated from run 2. $R_{\text{1G}} = 1$~G$\Omega (1+\delta R_{\text{1G}})$. (line, right axis): ULCA temperature output recorded during the calibration run. (d): Allan deviation of $1$~G$\Omega$ resistor value from (filled triangles) run 1, and (filled circles) run 2. Data obtained from a long run calibrating the same resistor using a CCC at $\pm 33$~nA is also shown for comparison (open circles). The dashed line with slope $1/ \sqrt{\tau}$ shows the expected Allan deviation due to the combined Johnson current noise in the $1$~G$\Omega$ resistor and the ULCA input stage.}}
\end{figure}

The method for calibrating a $1$~G$\Omega$ resistor using the ULCA has already been described in detail \cite{drung2015ultrastable} and is conceptually very simple. The resistor under test is used to convert a voltage from a voltage source into a current, and this current is converted back to a voltage using the ULCA in measurement mode. The difference between the ULCA output and the source voltage yields the unknown resistance in terms of the ULCA trans-resistance gain $A_{\text{TR}}$. The voltmeter measuring the difference voltage does not need to be accurately calibrated, but it does need good common-mode performance because the difference voltage is elevated above ground potential by the test voltage. In figure \ref{1GCalFig}, we show raw voltmeter data from two calibrations of the SET $1$~G$\Omega$. In both cases, the voltage source was switched between $\pm 4.8$~V. The data of panel (a) used the transfer ULCA, connected to the resistor in an adjacent laboratory using $\sim 10$~m coaxial cables. The data of panel (b) used the NPL ULCA placed adjacent to the  resistor and connected using short ($\sim 1$~m) cables, and the measured difference signal $\Delta V$ is indicated on the plot. The plots (a) and (b) share the same y-scale, and the increased noise using the long cables is evident. The resistor under test is evaluated as

\begin{equation}
R_{\text{1G}} = A_{\text{TR}} \Big( 1 + \frac{\Delta V}{\Delta V_{\text{source}}} \Big)
\end{equation}

with $\Delta V_{\text{source}} = 9.6$~V.

Panel (c), left axis, shows values of the resistor, expressed as deviation from nominal value in parts in $10^6$, obtained from a long measurement with the NPL ULCA  and short cables. The right axis shows the ULCA TEMP output during the measurement. This ULCA unit had temperature co-efficients, with respect to the TEMP output voltage, for $\delta G_{\text{I}}$ and $\delta R_{\text{IV}}$, of $0.033 \times 10^{-6}$~/~mV, and $-0.011 \times 10^{-6}$~/~mV respectively. The maximum excursion of the TEMP output, of $\approx 0.7$~mV, which corresponds to a temperature change of $\approx 0.36$~$^\circ$C, \footnote{For all the NPL measurements, the ULCA was in a temperature-controlled laboratory specified at $20.0 \pm 0.5 ^\circ$C} therefore changes $A_{\text{TR}}$ by less than $2$ parts in $10^8$. Similar stability was observed in all the long ULCA runs, for example the $G_{\text{I}}$ calibrations discussed in section \ref{GISec}. Immediately evident is a drift in the resistor value on time-scales of a few hours. This is quantitatively clear in the Allan deviation plot of the same data, figure \ref{1GCalFig} (d, filled circles) which shows a transition to $1/f$ noise on time-scales longer than about $1000$~s. Similar behavior, although with a larger noise background, was also seen using the long cables (filled triangles) and in a $1:100$ ratio measurement against a $10$~M$\Omega$ reference resistor using the CCC (also using the long cables) with $\pm 33$~nA flowing in the $1$~G$\Omega$ (open circles). The instability cannot be attributed to temperature fluctuations of the resistor, which had a measured temperature co-efficient of $5.2 \times 10^{-6} / ^{\circ}$C, and is temperature controlled with $\pm 0.005$~$^{\circ}$C stability. 

Similar instability has been observed now in several examples of commercially-made standard resistors, and it is believed to be an intrinsic property of the thick resistive film elements used to construct these standards \cite{giblin2018limitations}. The intrinsic noise of the $1$~G$\Omega$ ULCA measurement is mostly due to the thermal noise in the $1$~G$\Omega$ resistor itself, with a small additional contribution due to the thermal noise in the ULCA input stage. The Allan deviation expected from these two noise sources is shown as a dashed line in figure \ref{1GCalFig} (d). The noise in the ULCA output stage, $160$~nV$/ \sqrt{\text{Hz}}$, contributes a negligible amount to this measurement. It is interesting to compare the noise of the CCC and ULCA data. Using the long cables, the CCC achieves roughly a factor two lower relative uncertainty than the ULCA for short averaging times, but the excitation current is almost a factor 7 larger. The ability of the ULCA to resolve $\delta R_{\text{1G}}$ to better than $1 \times 10^{-7}$ with less than $5$~V across the resistor is useful in single-electron research, where small voltages $<1$~V are applied across the reference resistor, and the need to estimate the voltage co-efficient can increase the overall uncertainty \cite{giblin2012towards,zhao2017thermal}.

\begin{figure}
\includegraphics[width=8.5cm]{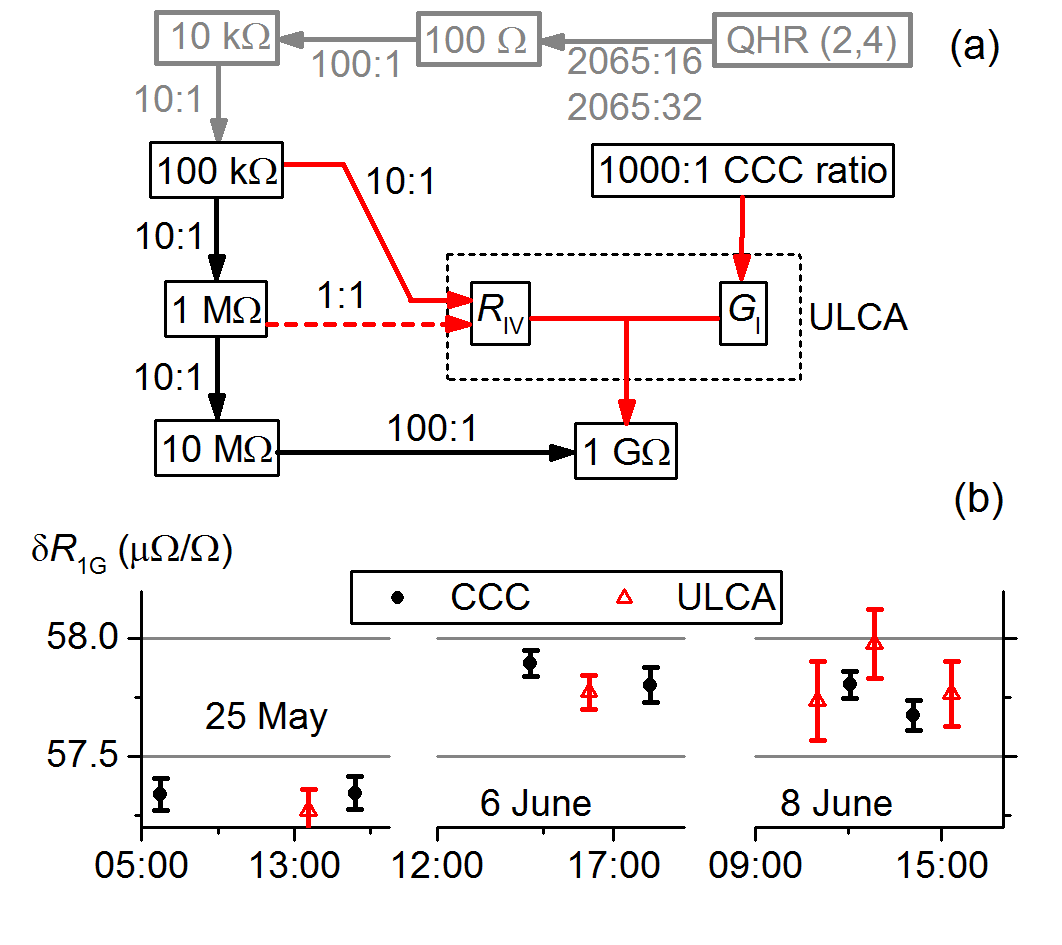}
\caption{\label{1GAgreeFig}\textsf{(a): Schematic diagram illustrating two routes for traceable calibration of a $1$~G$\Omega$ resistor: via CCC scaling (black arrows) and ULCA (red arrows). Additional grey arrows show traceability from $100$~k$\Omega$ back to the quantum Hall resistance (QHR) primary standard. (b): Measurements of a $1$~G$\Omega$ resistor using (filled circles) the CCC route, and (open triangles) the ULCA route. All x-axis date labels refer to the year 2017. $R_{\text{1G}} = 1$~G$\Omega (1+\delta R_{\text{1G}})$.}}
\end{figure}

Two routes for calibrating  $1$~G$\Omega$, via the ULCA (red arrows) and the established NPL CCC route (black arrows) are shown schematically in figure \ref{1GAgreeFig} (a). Both routes refer back to the same $100$~k$\Omega$ resistor (of Fluke type 742A) which was calibrated against the quantum Hall effect via two intermediate steps, indicated by grey arrows. We measured the SET $1$~G$\Omega$ via both routes on three different days, with the measurements taking place over a time span of not more than a few hours so that the data would not be affected by drift of the resistor described previously in this paper. This requirement was a severe constraint on the available measurement time, and in fact the smallest relative type A uncertainties achieved using the ULCA were $\approx 7 \times 10^{-8}$. In practice, it was was not necessary to perform the full set of measurements of figure \ref{1GAgreeFig} (a) every day; the $1$~M$\Omega$ resistor was sufficiently stable that the $100$~k$\Omega$ : $1$~M$\Omega$ measurement only needed to be performed every $\sim 10$ days. Likewise, the ULCA gain factors were stable to better than a part in $10^{7}$ over a few days. The results are presented in figure \ref{1GAgreeFig} (b), and show agreement at the level of $1-2$ parts in $10^{7}$. Larger error bars for the ULCA measurements on the last day were due simply to shorter measurement times.

\section{\label{ConcSec} Conclusions}

We have demonstrated that a small current source and measurement device, the ULCA, can be transferred between two laboratories by commercial courier with a change of its calibration factor of not more than $2$ parts in $10^7$, provided the temperature of the ULCA unit during the transfer does not drop below a critical threshold, which we tentatively identify as around $15 ^{\circ}$C. The trans-resistance gain $R_{\text{IV}}$ is more sensitive to transport effects than the current gain $G_{\text{I}}$. We speculate that this is because the former is determined by the absolute stability of a small number of resistor elements, while the latter is due to a network of thousands of individual, nominally identical resistors forming a matched resistor pair in which changes in the absolute parameters of the individual resistors do not strongly affect the overall resistance ratio (current gain). We have also demonstrated the calibration of the ULCA using a CCC originally designed for routine resistance calibrations long before the ULCA was conceived. The use of the ULCA to calibrate $1$~G$\Omega$ standard resistors has also been demonstrated, and we have shown that the ULCA has sufficient sensitivity and stability to resolve sub-ppm instabilities in these resistors. 

The stability of the ULCA at the sub-ppm level has profound consequences for small current metrology. It would firstly enable a proper verification of NMI small current calibration capability, without contaminating effects due the the stability of the transfer instrument. In fact, the stability of the ULCA has now been conclusively proven to exceed by at least an order of magnitude the smallest calibration uncertainties offered by NMIs in the field of small DC current. Secondly, the ULCA could also provide a step improvement in the way traceable small current measurements are performed in other fields, for example ionising radiation metrology. Here the requirement is for traceability at the $0.01 \%$ level, and in this setting an ULCA could function as a local current reference with a calibration interval of 5 years or even longer.

\begin{acknowledgments}
The authors would like to thank Eckart Pesel and Ulrich Becker for assistance with calibrations at PTB, and Colin Porter for assistance with calibrations at NPL. This research was supported by the UK department for Business, Energy and Industrial Strategy and the EMPIR Joint Research Project 'e-SI-Amp' (15SIB08). The European Metrology Programme for Innovation and Research (EMPIR) is co-financed by the Participating States and from the European Union's Horizon 2020 research and innovation programme.
\end{acknowledgments}

\bibliography{SPGrefs_ULCATransfer}

\end{document}